# New Investigations of Dark Floored Pits
# In the Volatile Ice of Sputnik Planitia on Pluto

S. Alan Stern
Southwest Research Institute

Brian Keeney
Southwest Research Institute

Rachael Hoover
Southwest Research Institute

Silvia Protopapa
Southwest Research Institute

Oliver White
SETI Institute

Will Grundy
Lowell Observatory

Dale P. Cruikshank
NASA Ames Research Center

And the New Horizons Team

Short Title: Sputnik Planitia's Dark Floored Pits
37 Pages
14 Figures
06 Tables

# Abstract

Sputnik Planitia, Pluto's gigantic, volatile ice glacier, hosts numerous scientific mysteries, including the presence of thousands of elongated pit structures. We examine various attributes of these pit structures in New Horizons datasets, revealing their length, aspect ratio, and orientation properties; we also study their reflectivities, colors, and compositions, and compare these attributes to some other relevant regions on Pluto. We then comment on origin mechanisms of the pits and also the fate of the missing volatiles represented by the pits on Sputnik Planitia. From a sample of 317 pits, we find typical length/width ratios of 2-4, with their major axis preferentially oriented approximately north-south. We also find that the floors of large pits in our sample have similar single scattering albedos and colors to dark material on crater rims and floors (i.e., possible subsurface windows) in Burney basin. We also find that the base of the three pits in our sample, large enough to study with LEISA IR spectroscopy, display both $CH_4$ and $N_2$ absorption features, as do the dark regions in crater windows in Burney basin. Evidence for a sublimation erosion origin for the pits is supported over both the explosion/ejecta venting and structural collapse alternatives. Finally, we find that the mass lost by the pits on Sputnik Planitia most likely lies condensed elsewhere, on Pluto's surface, relocated there by volatile transport as opposed to removal by escape to space or photochemical conversion.

# 1. Introduction

Among the most spectacular and scientifically exciting discoveries from the 2015 exploration of Pluto by NASA's New Horizons mission was the discovery of the massive $N_2$-$CH_4$-CO volatile ice glacier, Sputnik Planitia (henceforth SP), which dominates the flyby encounter hemisphere of Pluto (e.g., Stern et al. 2015; Moore & McKinnon 2021). SP is depicted in Figure 1.

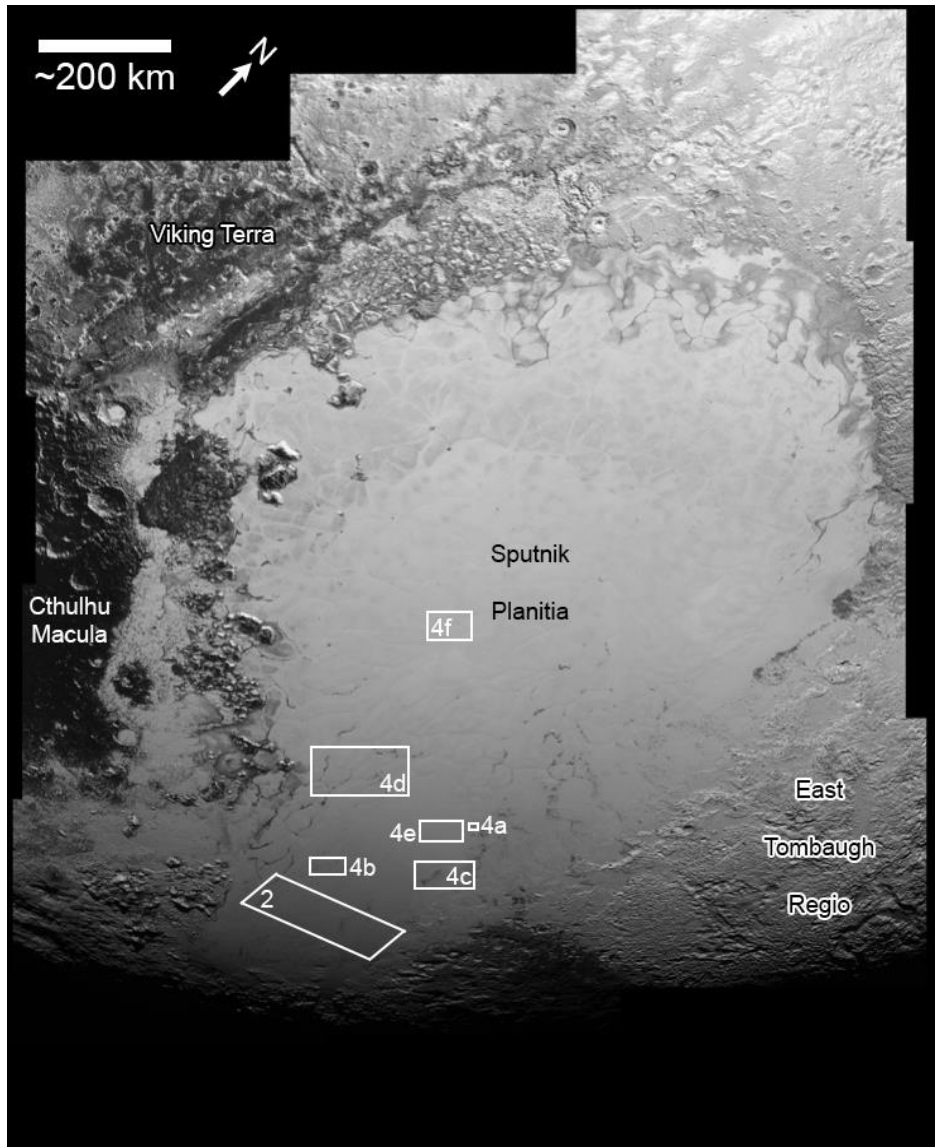

**Figure 1. Sputnik Planitia on Pluto. LORRI image mosaic obtained at 400 m pixel$^{-1}$, and centered at 174.4°E, 20.2°N. Locations of some surface features mentioned in the paper are indicated, as are the locations of Fig. 2 and panels a-f in Fig. 4, for context.**

SP's composition, geology, and geophysics have been explored and summarized in numerous publications (e.g., Moore et al. 2016; Grundy et al. 2016; White et al. 2017; Protopapa et al. 2017; Stern et al. 2018; Moore & McKinnon 2021).

In summary, as these and other reports have described, SP is a convecting ice sheet of ~800,000 km² extent, which has no discernable craters on its surface, indicating a crater retention age of <10 Myr (Moore et al. 2016; Singer et al. 2021). SP lies in an ~1300x1000 km² basin many km deep, of presumed impact origin created early in Pluto's history (e.g., Stern et al. 2015; McKinnon et al. 2016; Schenk et al. 2018; Moore & McKinnon 2021). As also described previously, there are multiple scientifically provocative aspects of SP, including evidence for convective overturning, lateral glacial dynamics, glacial recharge, buoyant $H_2O$-block mountains, and both dunes and wind streaks (Moore et al. 2016; Howard et al. 2017a,b; Umurhan et al. 2017; Telfer et al. 2018; Skjetne et al. 2021).

Here we concentrate on a study of the widespread dark floored pit features observed in SP. These structures (see Figure 2) have been discussed previously, notably in Moore et al. (2017), White et al. (2017), Buhler & Ingersoll (2018), and Wei et al. (2018). Fields of these pits correspond to the deeply pitted plains unit identified mainly in the southern and eastern areas of SP in the geomorphological mapping of White et al. (2017). Scipioni et al. (2021) calculated the average near-infrared spectra of geomorphological units mapped by White et al. (2017) for various provinces within SP, and determined that the floors of pits in this unit display the spectral signature of the "negative spectral slope end-member", i.e., a non-ice component presumed to be a macromolecular carbon-rich material, termed a tholin. This material has been suggested to either be lag deposit accumulations or exposed substrate under the bright surface ice (White et al. 2017; Moore et al. 2017).

In §2 here, we begin our study by examining 317 of SP's pits observed in the highest resolution New Horizons imaging of Pluto. For this selected sample, we measured pit dimensions and orientations. In §3, we discuss measured reflectivities and colors of this sample of dark floored pits and compare them to other dark exposed materials in windows into Pluto's subsurface seen on the New Horizons encounter hemisphere of Pluto. Then in §4 we examine the compositions of three large SP pits that are at least partially resolved by the New Horizons Ralph composition mapping spectrometer (Reuter et al. 2008). §5 provides a summary of the pit albedos, colors, and composition results. §6 then briefly discusses possible pit origin mechanisms, and §7 discusses the fate of the missing mass of volatiles lost from the dark floored pits. In §8 we summarize our overall findings.

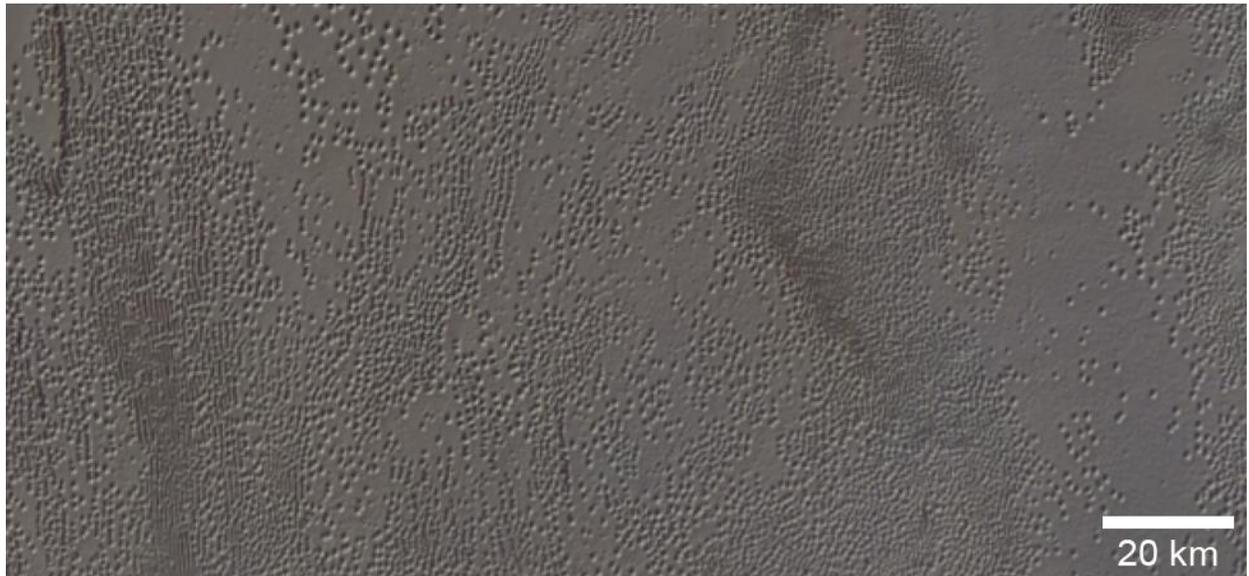

**Figure 2. A 315 m pixel$^{-1}$ New Horizons panchromatic image overlain by 650 m pixel$^{-1}$ color showing a pit field in southern Sputnik Planitia on Pluto. North is to the top; the image is centered at 185.4°E, 10.2°S.**

## 2. A Sampling of Pit Properties
## In High Resolution Sputnik Planitia Imaging

For this study, we cataloged 317 dark floored pits observed in the highest resolution (80-240 m pixel$^{-1}$) New Horizons Long Range Reconnaissance Imager (LORRI; Cheng et al. 2008) imaging of SP. All of these pits were selected to have lengths >1 km and widths >350 m, and all are resolved in the lowest resolution LORRI imaging used (240 m pixel$^{-1}$), which covers the entire study area. The locations of these cataloged pits are shown in Figure 3, and the properties of the images from which they were identified are summarized in Table 1.

The properties of the pits themselves are cataloged in Table 2. To produce the catalog, the outlines of the dark bottoms of the pits were traced in the geographic information system ArcGIS, which is maintained by the Environmental Systems Research Institute, and the centroids and areas of each pit were determined using the ArcGIS Tools for Graphics and Shapes extension. The length, width, and azimuths of the pits were estimated separately using a minimum-bounding rectangle to approximate the often-complex pit shapes. Since these are derived from a simplification of the shapes of the pits, the tabulated area is not equal to the product of the length and width.

The dark floored pits mostly occur in the non-cellular plains (White et al. 2017) of southern and western SP. They reach maximal dimensions >10 km (see Figure 4a), and can display elongate planforms that approach length to width ratios of 10:1 in extreme cases; they sometimes form doublets and aligned chains of joined pits (see Figure 4b). Additionally, they can cluster together in swarms that sometimes form arcuate, subparallel lines of pit chains (see Figure 4c). These pits are also seen to occur along cell boundaries and within expanses of non-cellular plains that are surrounded by the cells in central SP (see Figure 4d). The appearance of the dark floored pits is distinct from the smaller pits that are nearly ubiquitous across the non-cellular plains of SP (Figure 4e), as well as those in the cellular plains in the central and northern parts of SP (Figure 4f). These other pit forms only reach several hundred meters in diameter, and show only quasi-equidimensional planforms; they also do not exhibit dark floors.

Table 1: Properties of LORRI Images Used to Catalog SP Pits

| Identifier | Resolution [m/pix] | Latitude [deg] | Longitude [deg] | Area [km²] | Phase [deg] | Emission [deg] | Incidence [deg] |
|---|---|---|---|---|---|---|---|
| lor_0299177087_0x636 | 240 | 16.2 | 150.7 | 60,500 | 31.4 – 31.7 | 12.7 – 26.6 | 30.8 – 46.4 |
| lor_0299177123_0x636 | 240 | 10.7 | 156.1 | 60,400 | 31.4 – 31.7 | 13.5 – 30.3 | 38.0 – 53.4 |
| lor_0299177159_0x636 | 240 | 3.9 | 160.3 | 62,500 | 31.4 – 31.7 | 18.4 – 36.7 | 45.6 – 61.5 |
| lor_0299177195_0x636 | 240 | -4.5 | 164.9 | 67,000 | 31.3 – 31.6 | 26.0 – 45.7 | 54.4 – 71.6 |
| lor_0299177231_0x636 | 240 | -17.0 | 171.9 | 82,100 | 31.2 – 31.5 | 37.2 – 60.7 | 66.4 – 87.4 |
| lor_0299178911_0x630 | 120 | 10.1 | 174.3 | 14,700 | 47.8 – 48.1 | 3.9 – 12.2 | 51.9 – 58.0 |
| lor_0299178913_0x630 | 120 | 7.1 | 176.5 | 14,900 | 47.7 – 48.0 | 7.7 – 15.6 | 55.6 – 61.8 |
| lor_0299178915_0x630 | 120 | 4.0 | 178.7 | 15,100 | 47.5 – 47.8 | 11.6 – 19.2 | 59.3 – 65.6 |
| lor_0299178917_0x630 | 120 | 0.8 | 180.9 | 15,500 | 47.4 – 47.7 | 15.5 – 23.0 | 63.1 – 69.5 |
| lor_0299178919_0x630 | 120 | -2.5 | 183.2 | 16,000 | 47.2 – 47.5 | 19.5 – 27.0 | 67.0 – 73.5 |
| lor_0299178921_0x630 | 120 | -5.9 | 185.5 | 16,300 | 47.1 – 47.4 | 23.6 – 31.2 | 70.9 – 77.7 |
| lor_0299178923_0x630 | 120 | -9.4 | 187.8 | 17,200 | 46.9 – 47.2 | 27.8 – 35.6 | 75.0 – 82.0 |
| lor_0299179724_0x636 | 80 | 7.7 | 182.1 | 6,310 | 70.2 – 70.5 | 6.8 – 11.4 | 59.7 – 63.6 |
| lor_0299179730_0x636 | 80 | 3.8 | 184.6 | 6,200 | 70.1 – 70.4 | 2.1 – 7.1 | 64.3 – 68.2 |
| lor_0299179733_0x636 | 80 | 1.9 | 185.9 | 6,180 | 70.1 – 70.4 | 0.0 – 5.1 | 66.6 – 70.4 |
| lor_0299179736_0x636 | 80 | -0.1 | 187.1 | 6,190 | 70.0 – 70.3 | 0.0 – 4.3 | 68.9 – 72.7 |
| lor_0299179739_0x636 | 80 | -2.0 | 188.4 | 6,160 | 70.0 – 70.3 | 0.9 – 6.0 | 71.2 – 75.0 |
| lor_0299179742_0x636 | 80 | -3.9 | 189.6 | 6,170 | 69.9 – 70.2 | 3.2 – 8.0 | 73.4 – 77.3 |
| lor_0299179745_0x636 | 80 | -5.9 | 190.8 | 6,180 | 69.9 – 70.2 | 5.6 – 10.2 | 75.7 – 79.6 |
| lor_0299179748_0x636 | 80 | -7.8 | 192.1 | 6,210 | 69.8 – 70.1 | 7.9 – 12.5 | 78.0 – 81.9 |

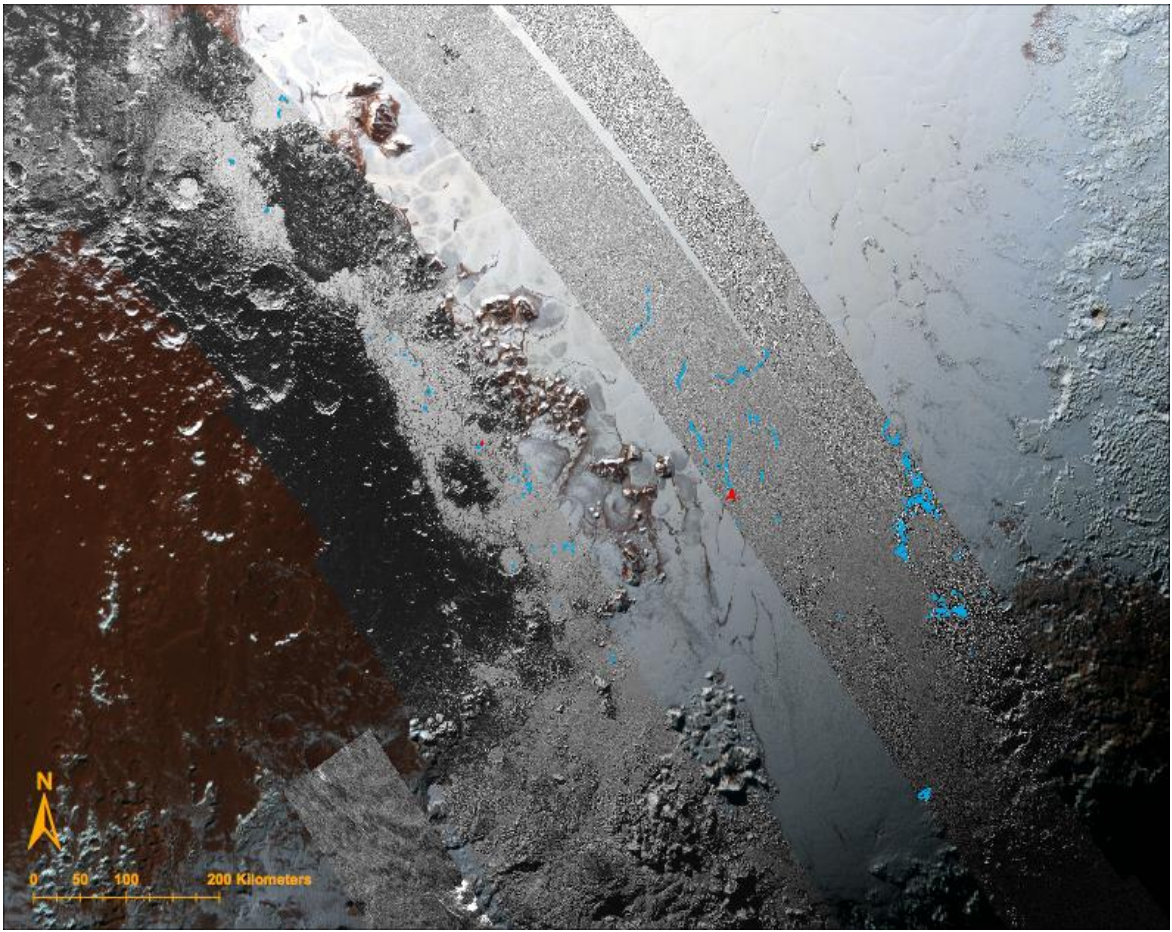

**Figure 3. LORRI high-resolution imaging (gray scale overlays); the red dots indicate the larger pits that were used for pit floor composition studies reported in §5. The base map here is the highest resolution New Horizons Ralph instrument Multi-color Visible Imaging Camera (MVIC) color image of Pluto's flyby encounter hemisphere, which has a resolution 3-8× coarser than the high-resolution LORRI images we used. The solar incidence angle is lowest in the northwest portion of the image, and highest in the southeast (see Table 1). Image is centered at 171.4°E, 1.2°N; north is to the top.**

Figure 3: Locations of pits in Sputnik Planitia (blue outlines) cataloged in

**Table 2: Catalog of Pit Properties**

| ID | Latitude [deg] | Longitude [deg] | Area [km²] | Length [km] | Width [km] | Azimuth [deg] |
|----|----------------|-----------------|------------|-------------|------------|---------------|
| 0  | 10.173 | 174.507 | 1.713  | 2.719  | 1.013 | 59.9  |
| 1  | 8.187  | 173.965 | 3.440  | 3.472  | 1.839 | 52.4  |
| 2  | 8.076  | 173.865 | 0.860  | 1.999  | 0.825 | 8.0   |
| 3  | 7.955  | 173.799 | 0.422  | 1.146  | 0.615 | -31.5 |
| 4  | 7.904  | 173.762 | 0.488  | 1.136  | 0.699 | -41.5 |
| 5  | 8.240  | 174.096 | 0.635  | 1.146  | 0.691 | -74.4 |
| 6  | 5.674  | 176.128 | 11.962 | 16.149 | 3.688 | 12.4  |
| 7  | 6.162  | 176.367 | 2.811  | 7.159  | 0.751 | 13.8  |
| 8  | 6.475  | 176.397 | 1.442  | 2.889  | 0.929 | 4.4   |
| 9  | 3.267  | 176.765 | 6.213  | 9.767  | 2.283 | -18.1 |
| 10 | 2.844  | 177.108 | 1.392  | 2.343  | 0.896 | 33.1  |

**Note:** Azimuth is the orientation of each pit's major axis, measured counterclockwise from north. Only a portion of this table is shown here to demonstrate its form and content. A machine-readable version of the full table was submitted and is available.

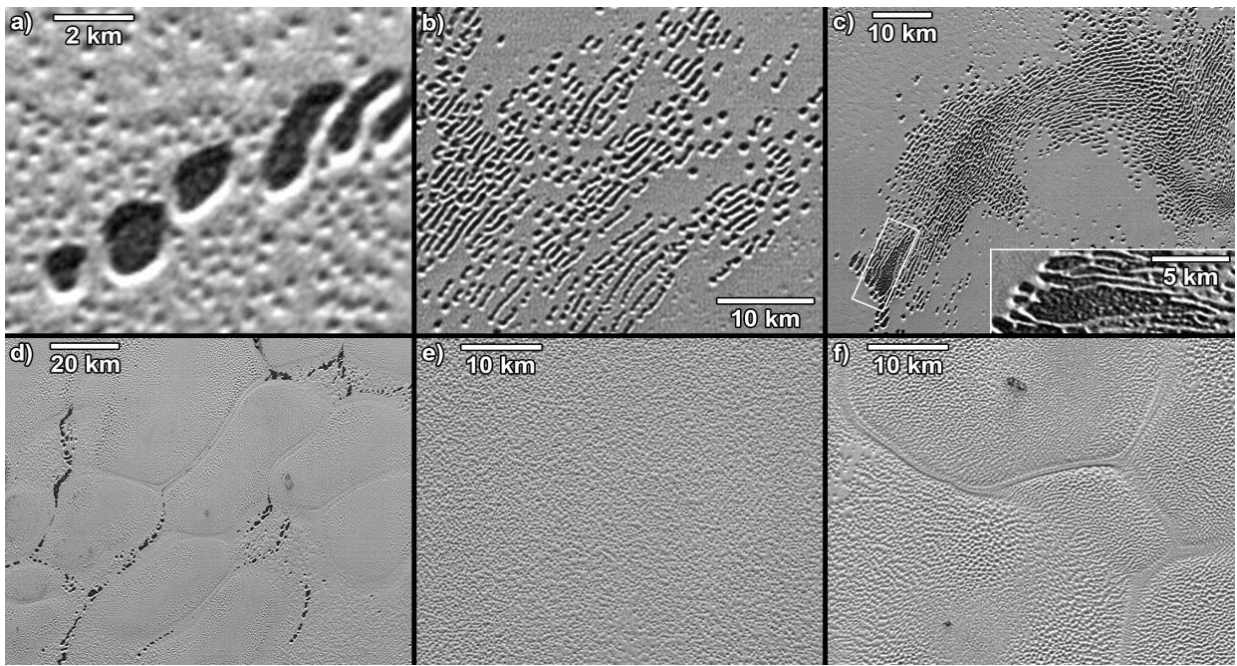

**Figure 4. Morphologies of pits across SP. North is to the upper right in all images, and illumination is from the top of each image. Image locations are indicated in Fig. 1. (a) Dark floored pits within the non-cellular plains that reach many km in diameter, and some of which exhibit elongate planforms. Shadow measurements (White et al. 2017) indicate that these pits typically attain depths between 160 and 190 m;**

76 m pixel$^{-1}$ LORRI image, centered at 186.9°E, 3.3°N. (b) A field of elongate and aligned pits within the non-cellular plains, in which isolated pits, doublets, and chains of joined pits can be seen; 117 m pixel$^{-1}$ LORRI image, centered at 182.8°E 5.3°S. (c) Densely spaced, elongate pits clustered together to form a wavelike swarm in the non-cellular plains. Dark floored pits reaching several kilometers across are seen at lower-left. The white box contains the largest pit of all in SP, measuring 12 km long by 2.5 km across. The pit is magnified in the inset, which shows that a rubbly texture at a scale of several hundred meters can be resolved on its floor; 76 m pixel$^{-1}$ LORRI image, centered at 188.3°E, 1.6°S. (d) Tendril-like clusters of dark floored pits located within narrow interstitials to convective cells of the cellular plains; 117 m pixel$^{-1}$ LORRI image, centered at 179.2°E, 3.5°N. (e) A dense field of non-dark floored pits, each reaching a few hundred meters across, in the non-cellular plains; 76 m pixel$^{-1}$ LORRI image, centered at 185.0°E, 1.5°N. (f) Non-dark floored pits within the cellular plains of central SP that reach maximum diameters of several hundred meters. The centers of the cells tend to be smooth, becoming more coarsely pitted toward the cell edges; 76 m pixel$^{-1}$ LORRI image, centered at 176.0°E, 16.5°N, shown at the same scale as Figure 4e.

**Figure 5: Some distributions of dark floored pit properties from our sample of 317 well-resolved pits. Panels (a)-(c) depict histograms of pit azimuths (measured clockwise from north), lengths, and length/width ratios, respectively. Panel (d) depicts pit length/width ratio as a function of pit length.**

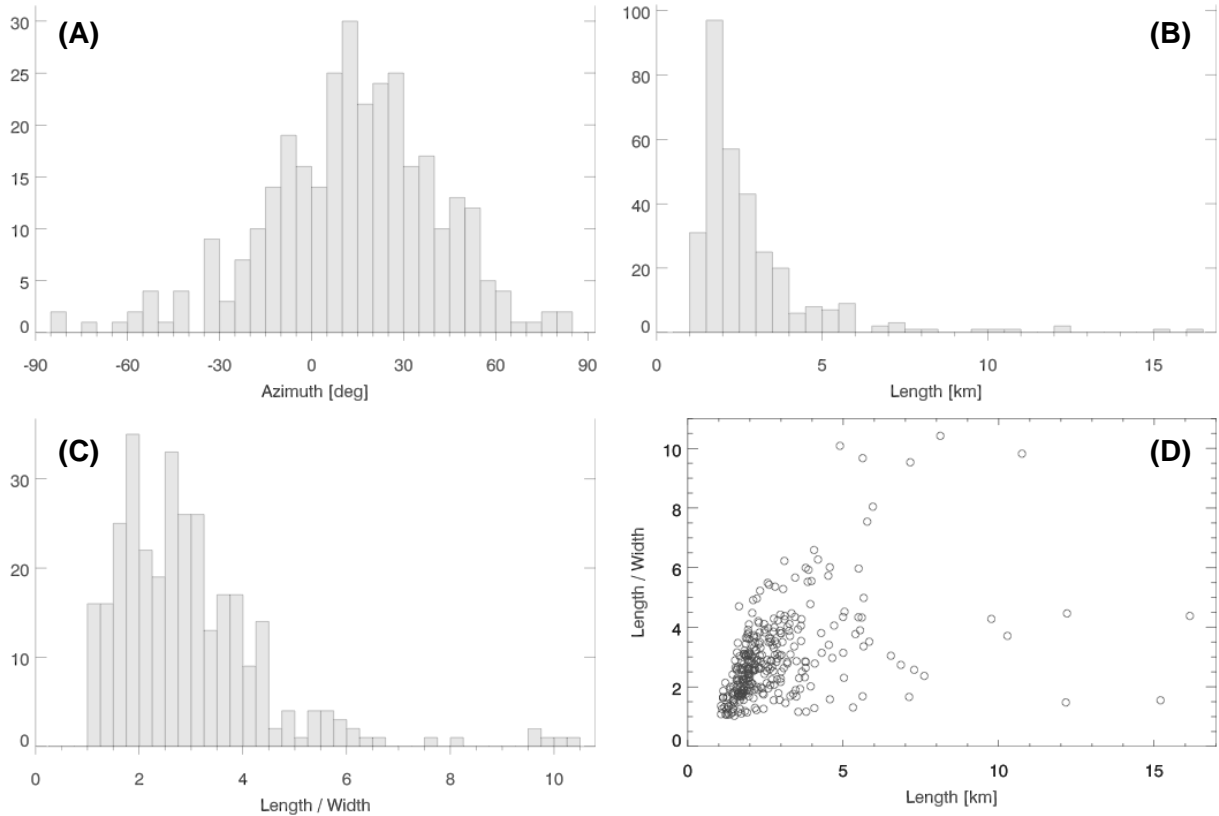

Figure 5 shows the distributions of several pit properties we measured in these 317 pits, including the azimuth of the major axis of these pits, measured from the north, their lengths, their length/width aspect ratios, and how pit length/width ratios correlate with pit lengths. From these results we infer that the typical pit in our catalog (i) is oriented approximately north to south, (ii) is 1-2 km in length, and (iii) displays an elongated length/width ratio of 2-4. Additionally, we find that the largest length pits (with lengths >10 km) do not display the extreme length/width ratios that some intermediate sized (i.e., 5-10 km length) pits do. Despite the substantial spatial spread of cataloged dark floored pits over the surface of SP (see Figure 1), we found no correlation of pit properties from our sample with either latitude or longitude (see Figure 6).

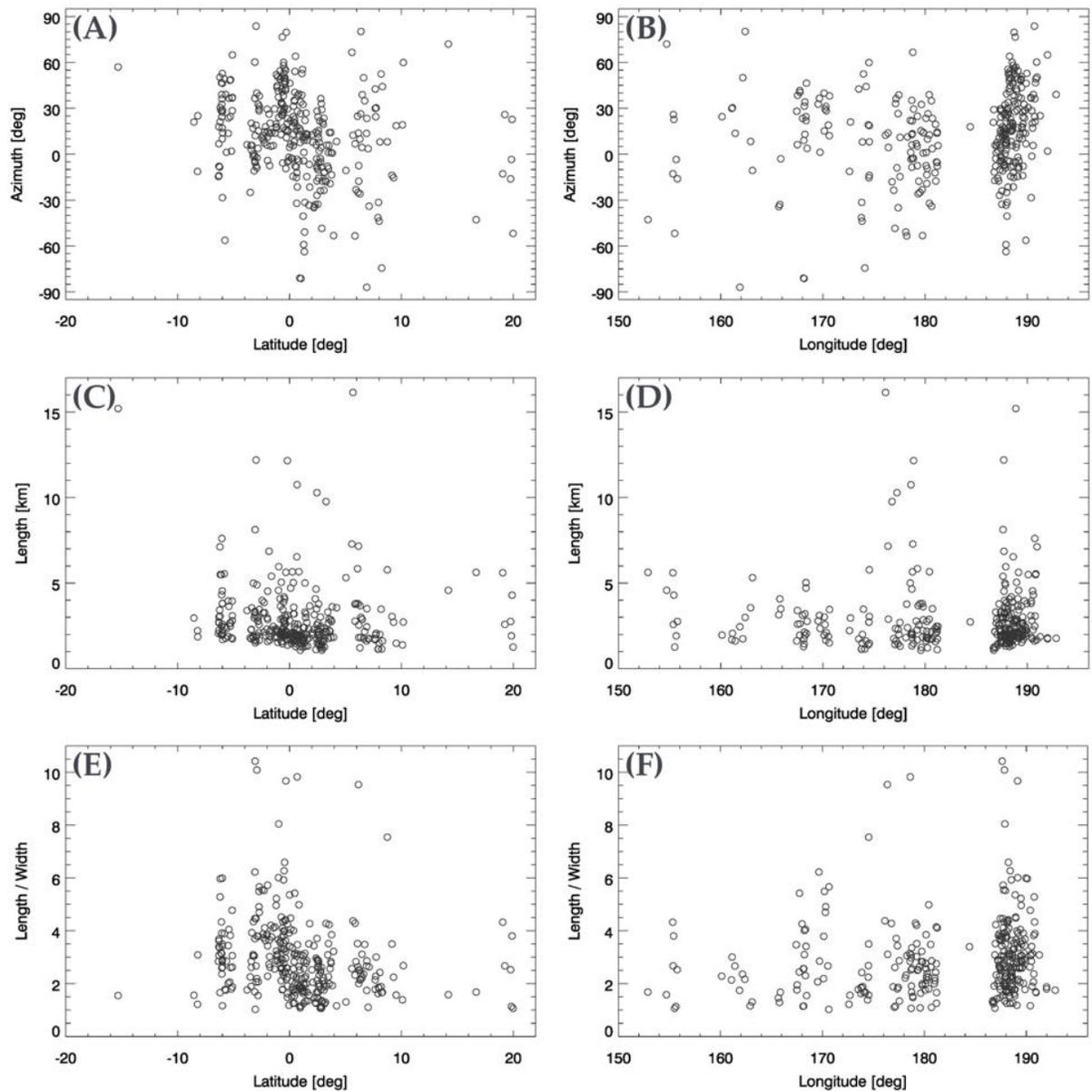

**Figure 6. Charts plotting three pit properties (azimuth, length, and length/width ratio) against pit latitude and longitude, showing that these dark floored pit properties in our sample do not correlate with longitude or latitude.**

Table 1 shows that the solar incidence angle varied systematically across the LORRI high resolution strips, causing more severe shadowing in the southernmost and easternmost portions of Figure 3. Because we are interested in the photometric properties of the dark floors of the pits, we excluded shadowed pixels from our analyses. We did this by assuming that all pits in our catalog had depths of 250 m, which is the depth of the largest pits in SP as derived from stereo topography (P. Schenk, personal communication); this is also broadly consistent with the pit depths ranging between 160 and 190 m as measured for a small number of pits using shadows (White et al. 2017). Using this conservative depth criterion and the specific Sun angle geometry for each pit in our catalog, we found that unshadowed pit floor pixels are observed on 245 (77%) of the pits in our catalog.

## 3. Sputnik Dark Floored Pit Colors and Albedos

We now explore the colors and single-scattering albedos[1] ($\omega$) of the 245 unshadowed SP dark floored pits described just above. In an effort to constrain the kind of material on the floors of SP's dark pits, we will then compare those color and albedo measurements to similar measurements on dark material in Pluto's Cthulhu Macula (Figure 1) and the dark, potentially formerly subsurface material seen exposed within impact craters in Pluto's Burney basin. We examine the Burney basin crater interior dark material on the hypothesis that they could have some commonality to the floors of SP's dark pits if the pit floors represent a widespread Pluto subsurface substrate. For the comparison with dark material in Burney basin craters, we chose exposed regions there that are comparable in terms of size and shape with the cataloged pits; the properties of the images containing these Burney Basin regions are listed in Table 3; these regions are depicted in Figure 7.

---

[1]All albedos quoted herein are single scattering albedos.

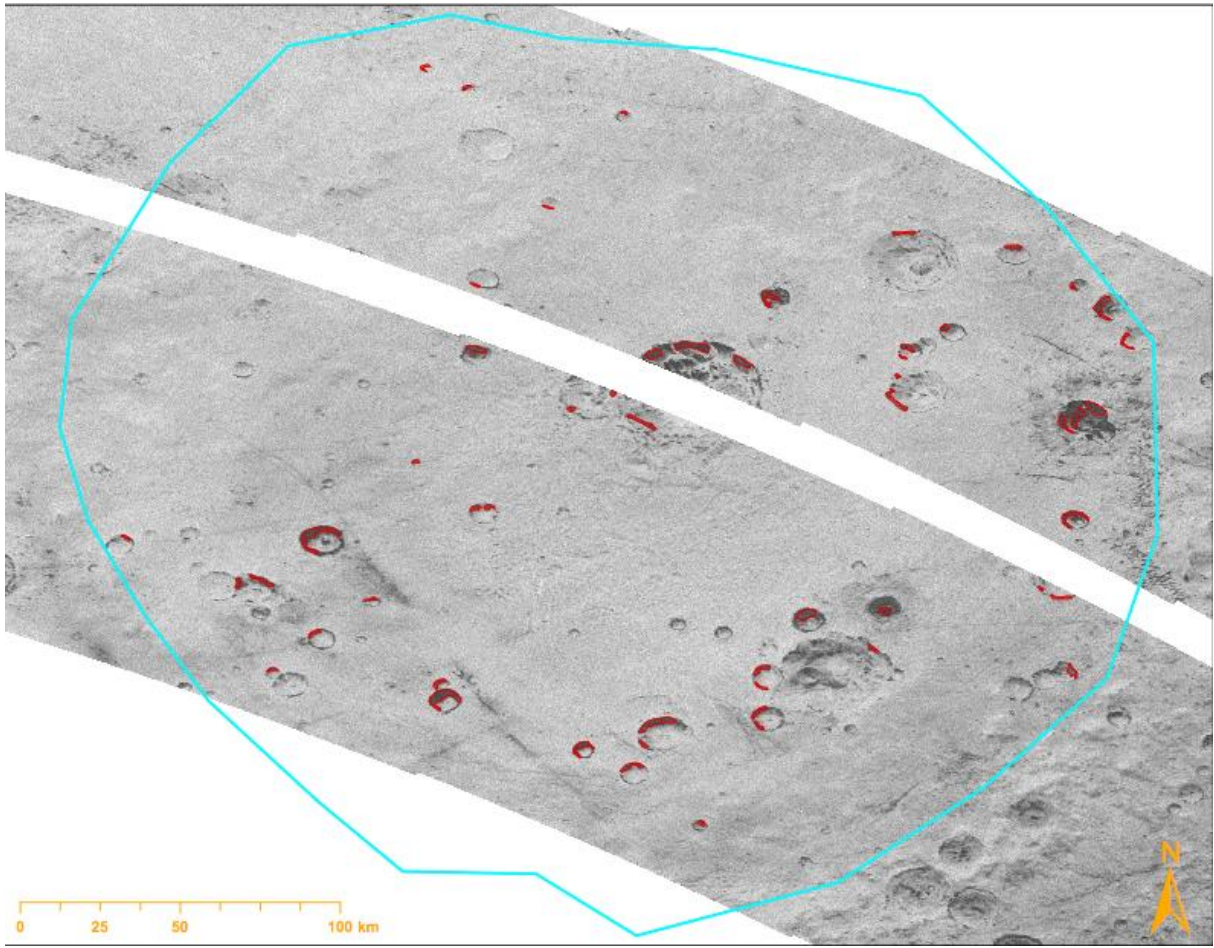

**Figure 7. Burney basin (cyan outline) and its dark crater material (red outlines) identified in high-resolution LORRI imaging (gray scale). North is to the top; the image is centered at 133.7°E, 46.0°N.**

Color information about Pluto's surface was measured by the MVIC color imager mentioned above; this imager has lower angular, and therefore lower spatial, resolution than the high resolution panchromatic LORRI imaging strips used to identify our sample of dark floored SP pits.

Despite this, the largest dark floored SP pits and the dark regions in Burney basin were resolved in the highest resolution MVIC color images at 660 m pixel$^{-1}$, allowing us to investigate their relative colors. Figure 8 is a color-color diagram of this dataset, designated P_Color2, showing the ratio of MVIC Red/Blue channel radiance factor (I/F) on the horizontal axis, and the ratio of MVIC NIR/Red channel radiance factor on the vertical axis. Contours are derived from all 16.3 million pixels in the P_Color2 dataset, which shows the entire encounter hemisphere of Pluto as observed by New Horizons; the median uncertainty is plotted in the upper left corner. The dense contours in the lower left of Figure 8 correspond to all of SP, and the contours in the upper right correspond to Cthulhu Macula.

Table 3: Properties of LORRI Images Used to Identify Dark Regions in Burney Basin

| Identifier | Resolution [m/pix] | Latitude [deg] | Longitude [deg] | Area [km²] | Phase [deg] | Emission [deg] | Incidence [deg] |
|---|---|---|---|---|---|---|---|
| lor_0299178885_0x630 | 120 | 46.1 | 127.4 | 23,200 | 49.7 – 50.0 | 43.7 – 54.0 | 1.7 – 9.4 |
| lor_0299178887_0x630 | 120 | 43.7 | 134.1 | 20,900 | 49.6 – 49.9 | 38.7 – 48.1 | 3.3 – 12.9 |
| lor_0299178889_0x630 | 120 | 41.1 | 139.8 | 19,300 | 49.4 – 49.7 | 34.0 – 42.8 | 8.1 – 16.8 |
| lor_0299179664_0x636 | 80 | 51.6 | 129.9 | 21,500 | 71.1 – 71.4 | 64.5 – 77.5 | 0.0 – 7.3 |
| lor_0299179667_0x636 | 80 | 48.5 | 138.3 | 16,300 | 71.1 – 71.4 | 59.6 – 69.5 | 2.0 – 11.8 |
| lor_0299179670_0x636 | 80 | 45.6 | 144.2 | 13,700 | 71.0 – 71.3 | 55.4 – 63.8 | 7.6 – 15.9 |

**Note:** The resolution, latitude, and longitude at the image center are tabulated, along with the full range of solar phase, emission, and incidence angles in the image.

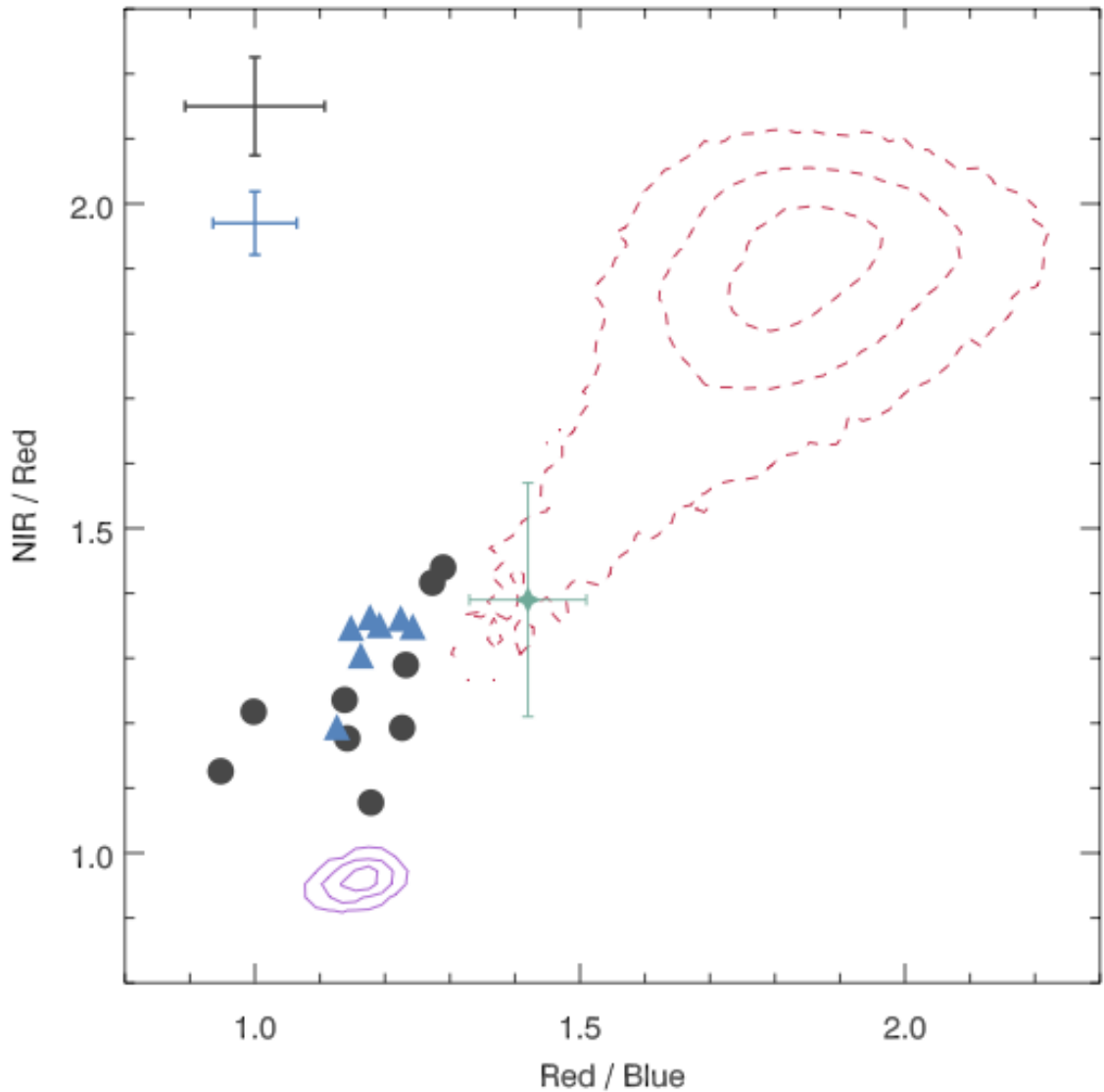

**Figure 8. Color-color diagram contours compared to the subset of SP dark pit floors large enough to be resolved in color datasets (circles), to resolved dark regions in crater walls in Burney basin (triangles), as well as to the entirety of SP and Cthulhu Macula. Redder colors plot in the upper right. The solid contours in the lower left correspond to SP; the dashed contours in the upper right correspond to Cthulhu. The median uncertainties of the SP pit floors and Burney dark regions are shown by the error bars in the upper left. Notice that the SP pit floors and dark regions in Burney have similar colors. The typical color of Viking Terra (point with error bar; see Table 6) is similar to, but distinct from, the colors of SP pit floors and dark regions in Burney.**

The filled dots in Figure 8 show the average colors of unshadowed pixels on the floors of SP pits; the triangles depict the average colors of the dark regions in Burney. Only SP pits and Burney dark regions that are fully resolved by MVIC are shown. These data reveal that SP pit floors and the dark regions in Burney have similar MVIC colors, suggesting that they may be compositionally similar, and potentially of similar origin. Further, these data also demonstrate that the colors of the SP pits and dark regions in Burney craters are both intermediate between those of SP and Cthulhu, which could imply that they are a mixture of bright SP-like volatiles and dark Cthulhu-like surface material.

More quantitative investigation requires employing a photometric model for Pluto. For this, we adopt the Hapke (2012) radiative transfer model implemented by Protopapa et al. (2020) to investigate disk-resolved photometric properties of Pluto. Protopapa et al. (2020) identified four Regions of Interest (ROIs) that are visible in five separate MVIC color images acquired at different viewing geometry (incident angle, emission angle, phase angle), and then solved for the single scattering albedo, $\omega$, and asymmetry factor, $\xi$, of each ROI as a function of wavelength in the range 492-883 nm. Note that $\xi$, can range from $-1$ to 1; the scattering is isotropic (p(g)=1) when $\xi$=0. If $\xi$>0, p(g) increases monotonically between 0 and $\pi$, and decreases monotonically if $\xi$<0. Given the limited solar phase angle coverage of the available data, Protopapa et al. (2020) considered the amplitude and width of the shadow-hiding opposition effect, as well as the macroscopic roughness, constant throughout the MVIC wavelength range and equal to the values obtained from a disk-integrated analysis of the Hubble Space Telescope data by Verbiscer et al. (2019).

Protopapa et al. (2020) found that to first order, $\xi$ is constant within the errors throughout the wavelength range, and equal to $-0.21\pm0.07$ (see their Table 3 and Figure 7). Our model is identical to theirs, except that we simplify it by setting $\xi$=$-0.21$ for all pixels. The contours in Figure 9 show the results of applying this simplified Hapke model to all pixels in the P_Color2 dataset. The horizontal axis here shows the spectral slope in % per 100 nm of each pixel at 550 nm; the vertical axis shows their single scattering albedo in the MVIC red channel centered at 624 nm. The purple, green, and red points here correspond to the values found by Protopapa et al. (2020) for SP, Viking Terra, and the dark equatorial region encompassing Cthulhu and Krun Macula, respectively. Our simplified model retrieves the correct albedo for all three ROIs, and the correct spectral slopes for the dark equatorial region. It does not retrieve the correct spectral slope for Cthulhu and Krun Macula, likely because these are the ROIs that showed the most variation of $\xi$ with wavelength in Protopapa et al. (2020). These results imply that our simplified photometric modeling of Pluto's surface derives reliable single scattering albedos, while our spectral slopes should be treated with caution.

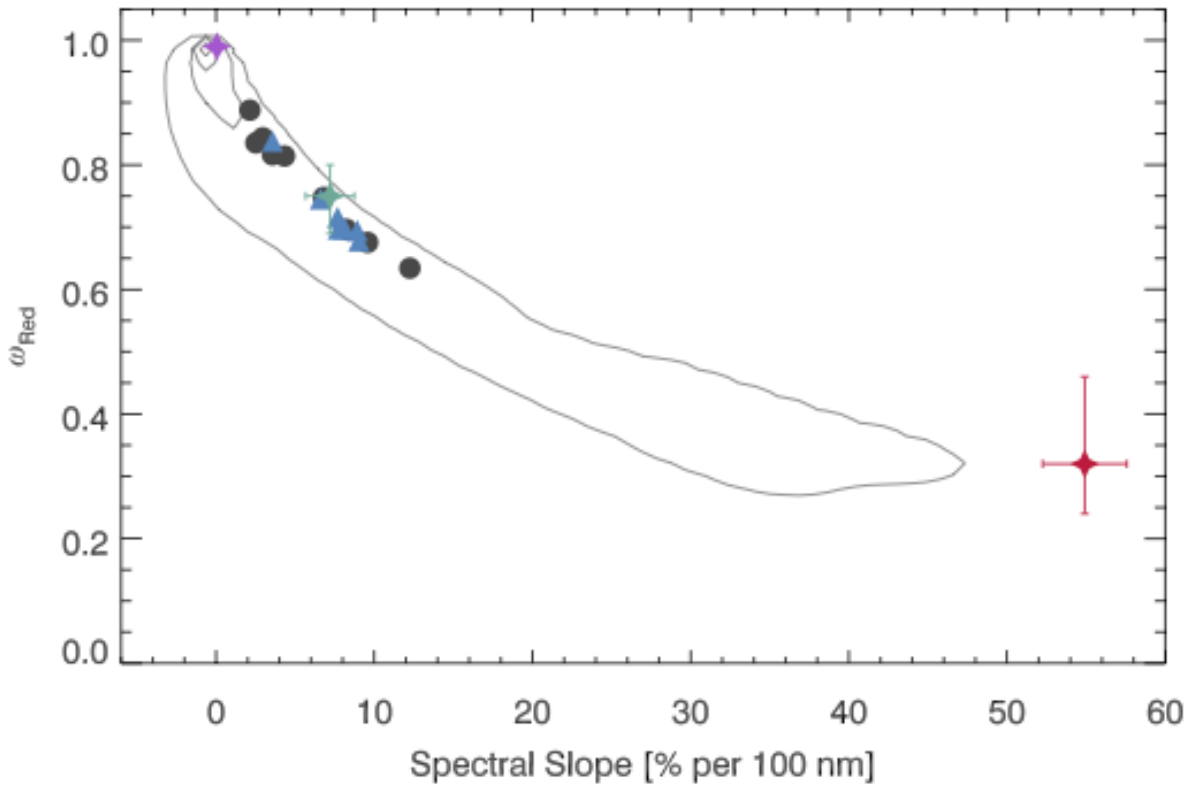

**Figure 9: Contour map of the number of pixels in a given MVIC Red-channel single scattering albedo/spectral slope bin, derived using the simplified photometric model described in the text. The purple, green, and red points correspond to the values found by Protopapa et al. (2020) for SP, Viking Terra, and the dark equatorial region, respectively. As in Figure 8, the circles and triangles are values for SP pit bottoms and Burney dark regions, respectively.**

After validating our modeling procedure this way, we applied it to the LORRI images (pivot wavelength 607nm; Buratti et al. 2017), leveraging their much higher spatial resolution relative to MVIC, to compare the single scattering albedos ($\omega$) of SP pit floors to the dark regions in Burney. When determining average single scattering albedo values, we restricted the sample to include only dark regions covering at least 10 contiguous pixels in area. This led to a final sample of 205 SP pit floors and 56 dark regions in Burney. However, since shadowing can split a single pit into several distinct un-shadowed regions, some of the SP pit floors in the sample probe the same pit; accounting for this, 187 unique pits are represented in the SP dark pit floor sample.

Histograms of both the average single scattering albedo distribution in our selected SP pit floors and the dark Burney regions are shown in Figure 10. The measured albedos of the SP Pits and dark Burney regions are listed in Tables 4 and 5, respectively. In both tables, N is the number of LORRI pixels in each delineated area, and the min and max albedo give the range of albedo calculated for the individual pixels. SP dark pit floors are seen to have an average single scattering albedo of $\omega=0.76\pm0.08$, slightly higher than the dark regions in Burney, which display an average $\omega=0.72\pm0.09$. However, the two distributions overlap significantly, and are not statistically different at high significance due to the small sample sizes. These average albedos are also both very similar to the single scattering albedo in the MVIC red channel of Pluto's Viking Terra ROI in Protopapa et al. (2020), and intermediate between the albedos of SP and Cthulhu (see Figure 9). Figure 10c shows the single scattering albedo for SP dark floored pits (circles) and Burney dark crater interior regions (triangles) as a function of area, and also shows consistency across the two regions. This panel suggests a trend of decreasing albedo with increasing area (i.e., larger regions of exposed substrate are darker), which may be caused by a smaller proportion of scattered light from pit walls being reflected by pit floors for the largest pits.

Table 4: LORRI Albedos of Pits

| ID | Pit ID | Area [km²] | Mean Albedo | Std Dev. | N | Min Albedo | Max Albedo |
|---|---|---|---|---|---|---|---|
| 0 | 1 | 1.356 | 0.715 | 0.059 | 98 | 0.374 | 0.884 |
| 1 | 3 | 0.199 | 0.735 | 0.032 | 14 | 0.639 | 0.845 |
| 2 | 6 | 4.918 | 0.669 | 0.058 | 346 | 0.423 | 0.979 |
| 3 | 7 | 0.453 | 0.718 | 0.047 | 30 | 0.605 | 0.820 |
| 4 | 8 | 0.818 | 0.727 | 0.059 | 56 | 0.602 | 0.862 |
| 5 | 9 | 0.760 | 0.723 | 0.089 | 53 | 0.570 | 0.908 |
| 6 | 10 | 0.372 | 0.775 | 0.087 | 24 | 0.562 | 0.907 |
| 7 | 11 | 0.371 | 0.761 | 0.078 | 24 | 0.531 | 0.902 |
| 8  | 12 | 0.233 | 0.749 | 0.077 | 16  | 0.638 | 0.884 |
| 9  | 13 | 5.938 | 0.672 | 0.089 | 410 | 0.422 | 0.967 |
| 10 | 14 | 0.572 | 0.791 | 0.065 | 40  | 0.650 | 0.912 |

**Note:** The second column lists the corresponding ID from Table 2; shadowing can split a single pit into multiple unshadowed regions. Only a portion of this table is shown here to demonstrate its form and content. A machine-readable version of the full table is available.

### Table 5: Properties of Burney Comparison Regions

| ID | Latitude [deg] | Longitude [deg] | Area [km²] | Length [km] | Width [km] | Azimuth [deg] | Mean Albedo | Std. Dev. | N    | Min Alb. | Max Alb. |
|----|---------------|----------------|-----------|-------------|------------|---------------|-------------|-----------|------|----------|----------|
| 0  | 47.909 | 131.725 | 5.695  | 5.961  | 1.897 | -89.9 | 0.634 | 0.035 | 598  | 0.409 | 0.869 |
| 1  | 45.128 | 129.307 | 18.145 | 12.251 | 7.951 | 71.5  | 0.574 | 0.037 | 1806 | 0.397 | 0.846 |
| 2  | 44.017 | 137.889 | 3.307  | 3.182  | 1.683 | -89.4 | 0.537 | 0.036 | 326  | 0.370 | 0.728 |
| 3  | 46.897 | 140.778 | 6.599  | 6.092  | 2.652 | 31.3  | 0.552 | 0.033 | 1513 | 0.405 | 0.681 |
| 4  | 48.680 | 136.135 | 6.632  | 5.651  | 4.440 | -60.4 | 0.603 | 0.044 | 1582 | 0.400 | 0.915 |
| 5  | 43.962 | 136.692 | 9.852  | 6.561  | 3.191 | 73.3  | 0.637 | 0.035 | 949  | 0.454 | 0.782 |
| 6  | 45.105 | 126.451 | 1.070  | 2.989  | 0.831 | -64.8 | 0.751 | 0.049 | 106  | 0.662 | 0.883 |
| 7  | 44.420 | 128.146 | 2.761  | 3.176  | 2.118 | -16.3 | 0.668 | 0.044 | 265  | 0.564 | 0.841 |
| 8  | 44.460 | 128.503 | 4.170  | 7.627  | 1.409 | -64.3 | 0.631 | 0.031 | 410  | 0.483 | 0.761 |
| 9  | 43.684 | 129.291 | 2.723  | 4.343  | 1.392 | 58.4  | 0.759 | 0.042 | 265  | 0.599 | 0.903 |
| 10 | 44.185 | 130.150 | 1.515  | 3.791  | 0.889 | 86.0  | 0.743 | 0.043 | 146  | 0.596 | 0.892 |
| 11 | 43.140 | 128.653 | 1.552  | 3.018  | 1.040 | 83.2  | 0.753 | 0.035 | 148  | 0.641 | 0.900 |
| 12 | 42.940 | 131.165 | 3.425  | 4.170  | 1.984 | 52.4  | 0.738 | 0.048 | 326  | 0.574 | 0.931 |
| 13 | 42.731 | 131.221 | 16.698 | 9.279  | 5.730 | 67.3  | 0.624 | 0.046 | 1578 | 0.433 | 0.855 |
| 14 | 45.546 | 131.717 | 2.428  | 3.628  | 1.449 | 73.7  | 0.781 | 0.041 | 244  | 0.638 | 0.928 |
| 15 | 45.560 | 131.923 | 2.035  | 3.111  | 1.409 | -87.9 | 0.786 | 0.040 | 207  | 0.598 | 0.933 |
| 16 | 41.982 | 133.310 | 7.507  | 6.149  | 3.880 | 54.7  | 0.663 | 0.072 | 699  | 0.512 | 0.926 |
| 17 | 41.693 | 134.055 | 5.352  | 7.591  | 2.007 | 76.6  | 0.735 | 0.074 | 499  | 0.549 | 0.945 |
| 18 | 42.346 | 134.443 | 13.431 | 12.022 | 2.939 | 72.2  | 0.655 | 0.076 | 1263 | 0.481 | 0.888 |
| 19 | 42.024 | 134.244 | 2.012  | 3.303  | 1.124 | -45.6 | 0.764 | 0.067 | 187  | 0.634 | 0.910 |
| 20 | 40.850 | 135.125 | 1.856  | 2.572  | 1.439 | -54.1 | 0.755 | 0.047 | 169  | 0.628 | 0.930 |
| 21 | 42.298 | 135.983 | 2.139  | 3.768  | 1.196 | -52.4 | 0.755 | 0.054 | 204  | 0.567 | 0.927 |
| 22 | 42.513 | 135.981 | 3.782  | 5.740  | 1.677 | 55.2  | 0.678 | 0.046 | 359  | 0.488 | 0.933 |
| 23 | 42.884 | 136.016 | 1.870  | 4.364  | 1.129 | -64.4 | 0.832 | 0.046 | 178  | 0.672 | 0.980 |
| 24 | 43.147 | 136.021 | 4.282  | 5.507  | 1.921 | 63.3  | 0.745 | 0.045 | 409  | 0.546 | 0.938 |
| 25 | 43.449 | 137.703 | 2.184  | 3.359  | 1.147 | -55.2 | 0.692 | 0.050 | 208  | 0.543 | 0.932 |
| 26 | 43.108 | 140.699 | 4.788  | 4.776  | 2.231 | -38.9 | 0.783 | 0.067 | 456  | 0.600 | 0.939 |
| 27 | 44.344 | 140.228 | 0.950  | 1.808  | 0.928 | -55.4 | 0.805 | 0.043 | 92   | 0.654 | 0.953 |
| 28 | 44.212 | 140.530 | 3.519  | 6.780  | 1.425 | -84.2 | 0.838 | 0.067 | 327  | 0.669 | 0.968 |
| 29 | 46.822 | 134.232 | 5.247  | 9.698  | 1.562 | -63.7 | 0.750 | 0.048 | 530  | 0.571 | 0.921 |
| 30 | 47.023 | 133.161 | 1.768  | 2.582  | 1.580 | 81.3  | 0.767 | 0.051 | 181  | 0.564 | 0.917 |
| 31 | 47.255 | 133.788 | 1.471  | 1.945  | 1.592 | 51.2  | 0.766 | 0.073 | 151  | 0.584 | 0.903 |
| 32 | 46.239 | 130.816 | 1.277  | 2.416  | 1.034 | 86.6  | 0.826 | 0.046 | 129  | 0.684 | 0.968 |
| 33 | 45.413 | 140.701 | 8.366  | 7.701  | 4.761 | 65.3  | 0.637 | 0.057 | 1863 | 0.431 | 0.965 |
| 34 | 48.125 | 141.517 | 1.006  | 1.898  | 1.221 | 82.0  | 0.770 | 0.055 | 239  | 0.600 | 0.943 |
| 35 | 47.999 | 141.494 | 2.133  | 5.115  | 2.056 | -41.6 | 0.831 | 0.054 | 502  | 0.583 | 0.981 |
| 36 | 48.541 | 141.143 | 9.348  | 7.198  | 5.291 | -1.7  | 0.635 | 0.061 | 2209 | 0.388 | 0.958 |
| 37 | 48.849 | 140.708 | 1.527  | 2.910  | 1.280 | -35.4 | 0.730 | 0.070 | 364  | 0.493 | 0.969 |
| 38 | 49.435 | 139.818 | 3.578  | 5.530  | 1.546 | -87.8 | 0.703 | 0.083 | 860  | 0.436 | 0.916 |
| 39 | 46.808 | 140.571 | 5.410  | 6.942  | 2.536 | -29.8 | 0.596 | 0.049 | 1240 | 0.405 | 0.881 |
| 40 | 47.027 | 141.048 | 10.056 | 6.945  | 3.102 | -56.5 | 0.600 | 0.040 | 2316 | 0.373 | 0.841 |
| 41 | 47.514 | 138.070 | 0.917  | 1.857  | 1.078 | -70.8 | 0.767 | 0.074 | 213  | 0.593 | 0.949 |
| 42 | 47.127 | 138.046 | 6.074  | 8.482  | 1.785 | -48.9 | 0.816 | 0.092 | 1396 | 0.588 | 0.988 |
| 43 | 47.299 | 137.987 | 0.586  | 1.453  | 0.786 | -48.5 | 0.692 | 0.058 | 136  | 0.593 | 0.907 |
| 44 | 47.935 | 138.221 | 3.703  | 4.672  | 1.904 | -69.6 | 0.680 | 0.079 | 863  | 0.501 | 0.901 |
| 45 | 47.794 | 138.150 | 0.937  | 2.997  | 0.778 | -67.0 | 0.828 | 0.079 | 220  | 0.649 | 0.976 |
| 46 | 48.234 | 138.780 | 2.423  | 2.390  | 1.937 | 49.4  | 0.664 | 0.068 | 569  | 0.508 | 0.915 |
| 47 | 49.648 | 138.174 | 3.158  | 7.199  | 1.083 | 88.9  | 0.771 | 0.043 | 763  | 0.614 | 0.959 |
| 48 | 51.441 | 133.954 | 0.794  | 2.207  | 0.855 | -76.0 | 0.811 | 0.048 | 198  | 0.658 | 0.961 |
| 49 | 50.029 | 132.806 | 1.066  | 3.748  | 0.554 | -71.5 | 0.885 | 0.050 | 264  | 0.715 | 0.994 |
| 50 | 48.870 | 131.714 | 1.035  | 3.363  | 0.708 | -63.9 | 0.789 | 0.064 | 248  | 0.649 | 0.980 |

| 51 | 47.841 | 134.403 | 7.017 | 5.346 | 2.961 | 73.8 | 0.615 | 0.068 | 1635 | 0.427 | 0.883 |
| 52 | 47.955 | 134.990 | 13.144 | 10.817 | 2.617 | -84.4 | 0.607 | 0.040 | 3081 | 0.434 | 0.861 |
| 53 | 47.725 | 135.731 | 6.190 | 6.076 | 2.035 | -66.4 | 0.677 | 0.039 | 1441 | 0.505 | 0.882 |
| 54 | 52.116 | 130.949 | 1.553 | 2.826 | 2.139 | 77.8 | 0.911 | 0.038 | 394 | 0.812 | 0.982 |
| 55 | 51.814 | 131.581 | 1.297 | 3.146 | 0.973 | 67.5 | 0.830 | 0.047 | 329 | 0.716 | 0.963 |

**Note:** Azimuth is the orientation of the region's major axis, measured counterclockwise from north.

SP dark pit floors and dark regions in Burney have essentially identical MVIC colors and single scattering albedos, which are considerably bluer and brighter than Cthulhu, and redder and darker than SP (see Table 6).

Dark regions in Burney basin were chosen for comparison to the SP pit floors because they are both plausibly windows into Pluto's subsurface. We find it striking how similar their photometric properties are to one another, but how dissimilar they are to other dark terrains (i.e., Cthulhu).

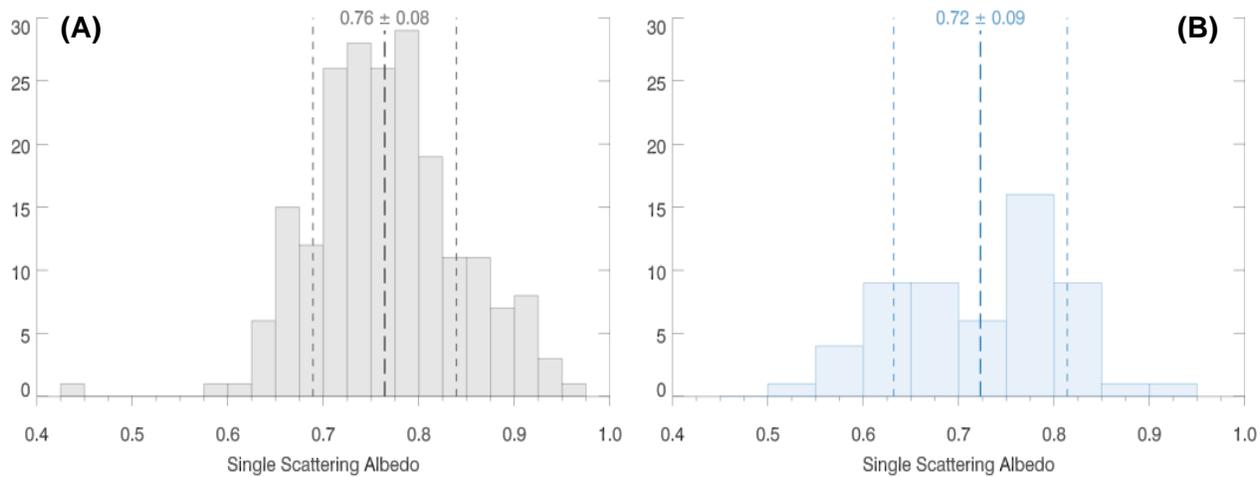

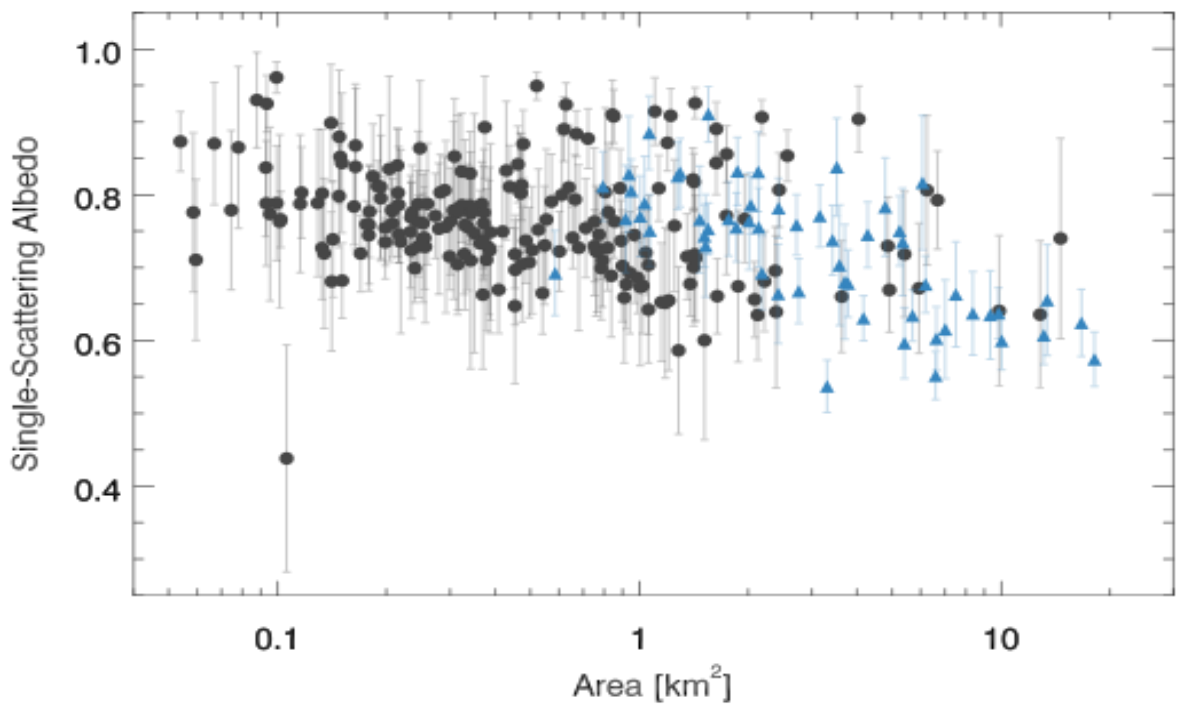

**Figure 10: Histograms of LORRI single scattering albedos for SP dark pit floors (panel A) and dark regions in Burney (panel B), with mean and standard deviation shown at their vertical dashed lines. Panel C shows the single scattering albedo for SP dark floored pits (circles) and Burney dark crater interior regions (triangles) as a function of area, and suggests a trend of decreasing albedo with increasing area (i.e., larger regions of exposed substrate are darker).**

## 4. Sputnik Dark Floored Pit Compositions

We utilized imaging from the New Horizons LEISA IR spectral imager (Reuter et al. 2008) to investigate the composition of the base of several large pits in SP. The locations of these pits are indicated by red dots in Figure 3. LEISA acquires data in the wavelength range of 1.25-2.50 µm, which contains unique absorption features for ices of $N_2$, $CH_4$, CO, and $H_2O$ (e.g., Grundy et al. 2016; Protopapa et al. 2017). We employed the highest spatial resolution LEISA imaging, which was obtained at mean solar phase angle of 33°, and has a mean image scale of 2.7 km pixel$^{-1}$. We identified three LEISA-resolved SP pits with both dimensions >3 km.

For each of these three large pits that the LEISA composition mapping spectrometer could resolve, we calculated an average spectral signature using pixels from within the pit for each pit base; see Figure 11. The spectral averages of these three pits were then compared to the average spectral signature of similar-sized areas of low albedo material within each of Cthulhu Macula, Viking Terra, and dark crater windows in Burney basin; see Figures 13 and 14.

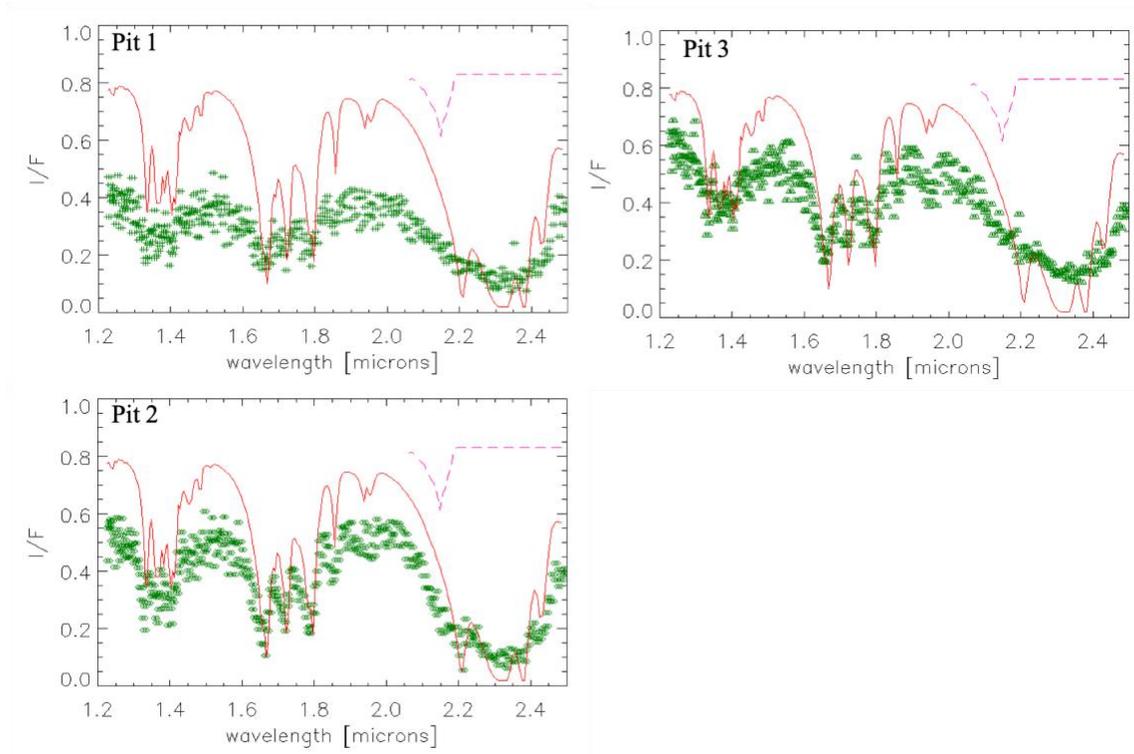

**Figure 11.** Average spectra (green symbols) for LEISA-resolved pits 1 , 2, and 3. For comparison, we show synthetic spectra for 100% $CH_4$ with particle diameter 500 μm (solid red line) and 100% $N_2$ with particle diameter 10,000 μm (dashed pink line) (Hapke 2012). A correction factor of 0.74±0.05 has been applied to the LEISA spectra per Protopapa et al. (2020). Notice that the three pit spectra are broadly similar, revealing $CH_4$ and $N_2$ absorption features in each. See Figure 12 which highlights the presence of the $N_2$ absorption feature.

As shown in Figure 11, the spectra for each of the three large pits have similar spectral signatures and all display $CH_4$ and $N_2$ absorption features. Because all pits present similar absorption features, we display the spectrum for pit 2 only in Figures 12-14. Given that pit 2 displays the strongest absorption features, for simplicity only pit 2 will be used for comparison to other regions on Pluto, though comparing to pits 1 or 3 would lead to the same conclusions. To better discern the $CH_4$ and $N_2$ absorption features, we show the spectrum of pit 2 compared to Hapke's (2012) radiative transfer synthetic spectra of $CH_4$ and $N_2$ as well as to the ratio of the pit spectrum to the synthetic $CH_4$ spectrum in Figure 12. We also point out that based on thermodynamic equilibrium considerations, pure $CH_4$ is not expected to coexist with pure $N_2$ on Pluto's surface. Where both species are present on the surface of Pluto, two phases occur (Trafton 2015; Protopapa et al. 2015, 2017): $CH_4$ saturated with $N_2$, and $N_2$ saturated with $CH_4$.

$N_2$ and $CH_4$ on pit bottoms may be a simple thin layer of a seasonal or secular condensate over the lower albedo material below. In this scenario, that darker material is likely a lag deposit that is exposed by the sublimation of the ice. When tiny dark particles are entrained in an otherwise fairly transparent crystal, only a very small concentration is needed because the coloring effects are greatly enhanced, as incident photons have a long path length in transparent volatile ice grains.

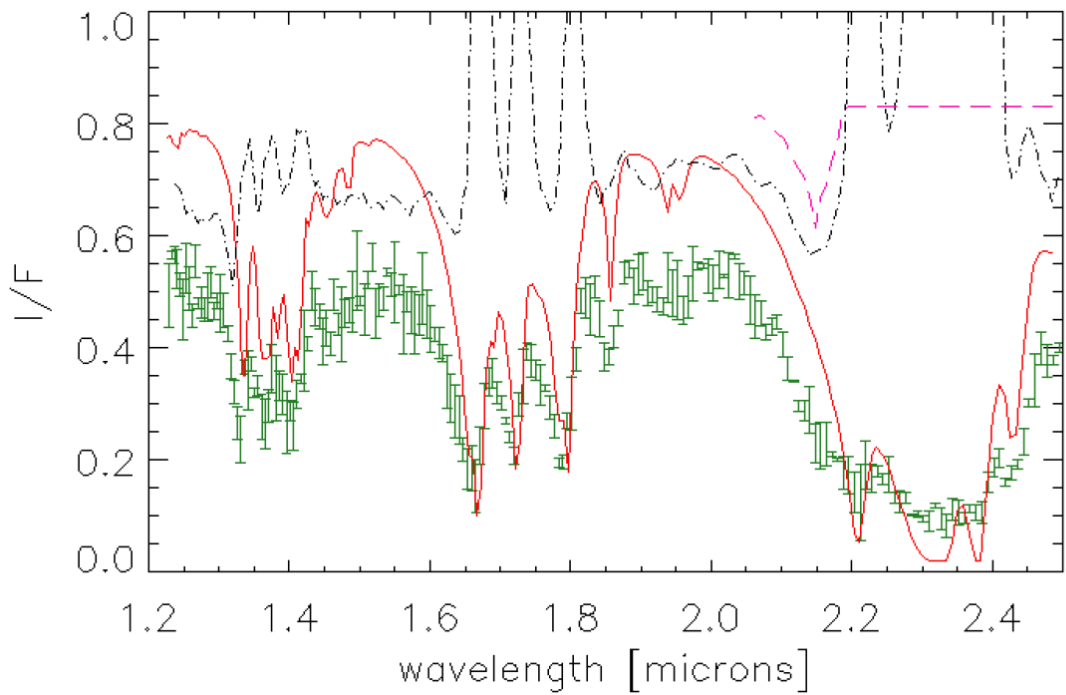

**Figure 12. Average spectrum of the base of pit 2 (green bars), compared to the synthetic spectrum (Hapke 2012) for $CH_4$ with particle diameter 500 μm (solid red line), and a synthetic spectrum for $N_2$ with particle diameter of 10,000 μm (dashed pink line). The error bars represent one standard deviation of the spectrum. Also plotted is the ratio of the pit 2 spectrum to that of synthetic $CH_4$ (dash-dot black line). This ratio more clearly shows the presence of the $N_2$ absorption band around 2.15 μm. The base of pit 2, like pits 1 and 3 (Figure 11), shows $CH_4$ and $N_2$ absorption features. A correction factor in I/F of 0.74±0.05 (Protopapa et al. 2020) has been applied to all of the LEISA spectra.**

We also found that the spectra of the bases of the pits are qualitatively similar to the dark material in Burney basin (Figure 14), which is dominated by CH$_4$ absorption features, but are spectrally distinct from both Cthulhu Macula and Viking Terra (Figure 13), which instead display clear evidence of H$_2$O ice absorption bands that are not seen in the pit floors large enough for LEISA to resolve.

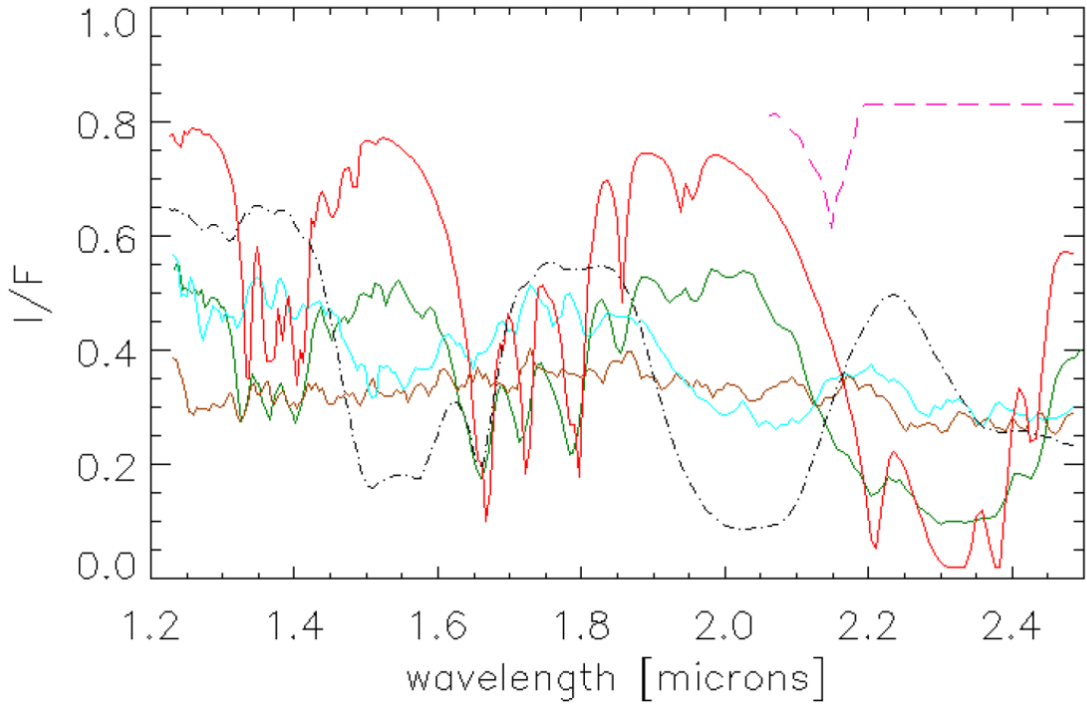

**Figure 13. Comparison of the spectrum of the base of pit 2 (green) to the spectra of Cthulhu (brown) and Viking Terra (blue), both of which are spectrally different from the pit. These are compared to the synthetic spectra (Hapke 2012) of CH$_4$ (solid red line), N$_2$ (dashed pink line) and H$_2$O (dash-dotted black line) with a particle diameter of 500 μm, 10,000 μm, and 100 μm, respectively. A correction factor of 0.74±0.05 in I/F (Protopapa et al. 2020) has been applied to all of the LEISA spectra.**

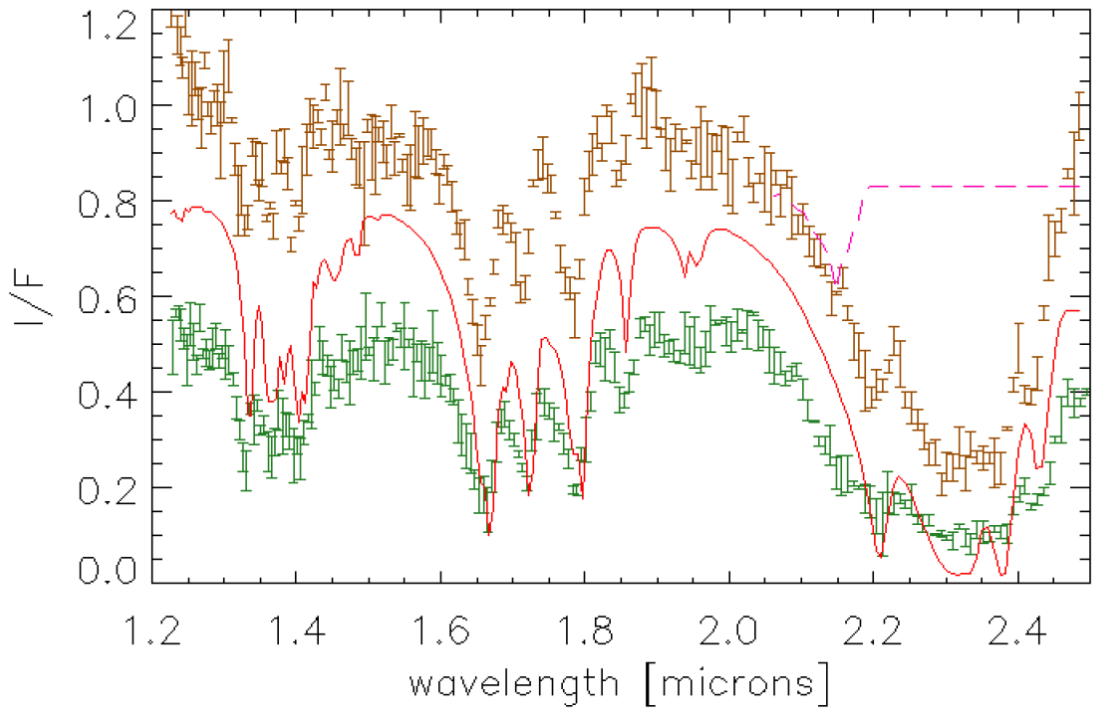

**Figure 14.** Comparison of the dark material in Burney basin (brown) to the base of pit 2 (green) shows they are spectrally similar and dominated by $CH_4$ absorption. The error bars represent one standard deviation of the spectrum. These are compared to the synthetic spectra (Hapke 2012) of $CH_4$ (solid red line) and $N_2$ (dashed pink line) with a diameter of 500 μm and 10,000 μm, respectively. A correction factor in I/F of 0.74±0.05 (Protopapa et al. 2020) has been applied to all of the LEISA spectra.

Finally, the spectral properties of these three large dark floored pits were also compared to plains material on Sputnik Planitia itself, using pixels adjacent to each of the three pits. This was done to discern whether the spectra observed within the bases of the pits is representative of the true pit floor composition or if it might be substantially contaminated due to either pixel smear or optical resolution limitations of the LEISA data. Tests we performed cannot explain the $N_2$ and $CH_4$ detected in these three large pits purely as a result of optical mixing from adjacent SP plains pixels.

## 5. Summary of Sputnik Pit Albedo, Color, and Composition Results

In Table 6, we summarize our comparison of single scattering albedos, color ratios, and compositions for the measured SP pits to the average of all SP terrains, and to the regions of interest in the Burney dark material, Cthulhu Macula, and Viking Terra.

**Table 6: Summary of Measured Sputnik Planitia Pit Reflectance and Compositional Properties**

| Area | ω | Red/Blue Color Ratio | NIR/Red Color Ratio | Composition |
|---|---|---|---|---|
| Burney Basin Dark Material | 0.72±0.09 | 1.18±0.04 | 1.35±0.02 | $CH_4$ and $N_2$ |
| SP Dark Pits | 0.76±0.08 | 1.16±0.13 | 1.24±0.14 | $CH_4$ and $N_2$ |
| SP Average | 0.99±0.01 | 1.10±0.05 | 0.95±0.04 | $CH_4$ and $N_2$ |
|  |  |  |  |  |
| Cthulhu Macula | 0.32±0.11 | 1.80±0.12 | 1.87±0.12 | Lacks $CH_4$, $N_2$ |
| Viking Terra | 0.75±0.06 | 1.42±0.09 | 1.39±0.18 | Lacks $CH_4$, $N_2$ |

Comparing the measured SP pits to SP as a whole, we find that the pits are (i) considerably less reflective, (ii) almost indistinguishable within the uncertainties in their Red/Blue color ratio, (iii) have a redder NIR/Red ratio, and (iv) are broadly similar in spectral behavior in the near-IR.

Comparing the SP pits to the three other dark regions of interest, we find striking single scattering albedo, color, and compositional similarities to the dark material in Burney basin windows, but distinct differences from Cthulhu Macula's (albedo) and Viking Terra's (color) . Why the bases of the SP pits and the subsurface material exposed in Burney basin should be so similar, despite their widely separated locales and latitude, is unclear, but may suggest a widespread uniformity of exposed subsurface materials beneath volatile, covered units across Pluto.

## 6. On the Fate of the Missing Mass from the Pits

Finally, we want to consider the "missing mass" of material represented by the present day dark floored pits on SP. Using LORRI imaging and pit coverage fractions from Buhler & Ingersoll (2018), we derived a total dark floored pit surface area on SP of 82,000 km². As described above, topographic data retrieved from photoclinometric digital terrain models created for SP (Schenk et al. 2018; Beyer et al. 2019) indicate a characteristic pit depth of 100 m. Combining these numbers and converting to cgs units, we estimate a total dark floored pit volume of $8.2 \times 10^{18}$ cm³, which corresponds to a total missing mass for zero porosity ice equal to $8 \times 10^{18}$ gm for $N_2$ ice and $4 \times 10^{18}$ gm for $CH_4$ ice. Even if SP's surface ices are highly porous, it is hard to imagine there is $<1 \times 10^{18}$ gm of missing material represented by all of SP's dark floored pits.

By comparison, the mass of Pluto's present day atmospheric $N_2$ and $CH_4$ inventories are just $3 \times 10^{16}$ gm and $1 \times 10^{14}$ gm, respectively (Young et al. 2018). This atmospheric inventory is smaller by orders of magnitude than the missing mass represented by the present day pits. This, in turn, leads one to conclude that the missing mass of volatiles represented by the dark floored pits cannot be resident in Pluto's atmosphere today, and therefore, must either have been removed by escape or photochemical destruction, or atmospherically transported and re-condensed elsewhere on the planet.

In what follows in this section we therefore explore how $N_2$ removed from SP's pits may have escaped as a general part of Pluto's well known and measured atmospheric escape (Gladstone et al. 2016) and/or (as the two processes are not mutually exclusive) may have been transported through the atmosphere to now lie elsewhere on the planet.

Regarding the atmospheric escape scenario, Young et al. (2018) derived escape rate estimates for $N_2$ and $CH_4$ of $5 \times 10^{22}$ molecules s$^{-1}$ and $6 \times 10^{25}$ molecules s$^{-1}$, respectively. Converting these numbers to mass escape rates using $N_2$ and $CH_4$ molecular weights in turn yields characteristic escape timescales for the missing $N_2$ and $CH_4$ mass, of $4 \times 10^9$ and $3 \times 10^7$ years, respectively. Regarding the photochemical destruction of SP dark floored pit volatiles by UV sunlight, Krasnopolsky (2020) gives photochemical destruction rates of $N_2$ and $CH_4$ of $\sim 7 \times 10^9$ gm yr$^{-1}$ and $3 \times 10^2$ gm yr$^{-1}$, respectively. From this we derive $N_2$ and $CH_4$ pit mass loss timescales of $1 \times 10^9$ and $4 \times 10^7$ yrs, respectively.

All of these timescales are long compared to Pluto's 248-year orbital-seasonal cycle and to its 3.8 Myr mega-seasonal obliquity cycle (e.g., Binzel et al. 2017). More importantly, they are long compared to the ~1

Myr retention age of km-sized pits forming within a shallow layer of enhanced viscosity $N_2$ ice as determined by Wei et al. (2018). This indicates that the missing mass of SP's pits cannot have been removed by escape or chemical processes since their formation, and likely must be removed by volatile transport.

This again suggests that the missing pit mass lies condensed elsewhere on Pluto's surface, relocated there by volatile transport through the atmosphere. For reference, the minimum present day mass loss of dark floored pit volatiles derived above, $\sim 10^{18}$-$10^{19}$ gm, represents an average layer of several tens of gm cm$^{-2}$ spread across Pluto's surface, depending on porosity and $N_2$ or $CH_4$ composition of the sublimated dark floored pit ices. This is an order of magnitude smaller than the estimated total volatile transport inventory migrating with Pluto's obliquity cycles (Bertrand et al. 2018), which is also consistent with the volatile transport mechanism.

Large deposits of $N_2$ ice exterior to SP are seen within Pluto's equatorial regions, most notably the bright, pitted uplands of East Tombaugh Regio (Figure 1), where $N_2$ ice appears to have condensed across the entire rugged landscape and ponded in depressions (Howard et al. 2017a,b; White et al. 2017), as well as in low-elevation basins and craters both to the west of SP and to the east of Eastern Tombaugh Regio (Stern et al. 2021). In addition, $N_2$ ice is present at the surface across much of the high northern latitudes (Grundy et al. 2016; Protopapa et al. 2017; Schmitt et al. 2017), albeit as a veneer that does not form large-scale deposits like those at the equator. These and other $N_2$ deposits may, at least in part, represent SP dark pit mass loss volatiles.

### 7. Regarding the Origin of Sputnik's Dark Floored Pits

The origin of the pits seen on SP has been discussed since their discovery (e.g., Stern et al. 2015; Grundy et al. 2016; Moore et al. 2016; White et al. 2017; Moore et al. 2017).

The most commonly proposed origin mechanism for SP's dark-floored pits is sublimation erosion (Moore et al. 2017; White et al. 2017), a mechanism that has also been discussed for other kinds of pits on Pluto's surface (e.g., Howard et al. 2017b). SP's dark-floored pits reach much larger diameters than any pits seen in SP's cellular plains, which has been attributed to the fact that on non-cellular plains and along the margins of convection cells, the absence of convective overturning allows pits to grow to much larger sizes before they are eventually destroyed by viscous flow of the ice (Buhler & Ingersoll 2018; Wei et al. 2018). This interpretation would, in turn, seem to imply that SP's dark floored pits are older than the smaller pits on both the cellular and non-cellular plains, though as we discuss below, relating pit size to pit age is dependent on the rate at which SP's dark floored pits grow relative to SP's smaller pits, which will depend on the rheological properties and perhaps other active processes besides sublimation in the non-cellular plains (compared to the cellular plains).

An important consideration regarding pit growth is what rheology of the nitrogen ice is necessary to allow it to support the higher pit walls of the dark floored pits. Sublimation-induced pitting appears to compete with diffusive flow of the near-surface ices, which tends to smooth the surface. Moore et al. (2017) and Wei et al. (2018) both found that the retention age for a 1 km diameter pit within $N_2$ ice with the viscosity measured in the lab by Yamashita et al. (2010) is <1 terrestrial year. This is inconsistent with the observed multi-km diameters and >100 m depths of the dark floored pits, as sublimation rates of hundreds of meters per Pluto year would be necessary to offset the rapid infilling of the pits via viscous flow of the $N_2$ ice, which are far in excess of what has been modeled for southern Sputnik Planitia (<18 mm per Pluto year, see Bertrand et al. 2018). Wei et al. (2018) argued that a layer of enhanced viscosity reaching a few hundred meters thick is necessary to support the observed topography of pits at the scale of the large, dark floored pits. The results of their numerical modeling show that for a 200 m thick surface layer with a viscosity $10^8$ times the experimental value of Yamashita et al. (2010), overlying an interior layer with a viscosity $10^4$ times the experimental value, the retention age of a 1 km diameter, 200 m deep pit is >$6 \times 10^5$ years. Wei et al. (2018) cited grain size growth through sintering potentially higher concentrations of $CH_4$ and CO, and the possibility that the surface layer of $N_2$ ice falls into a different creep regime, as ways to create such a layer. Based on this revised retention age of Wei et al. (2018), for a 1 km diameter, 200 m deep pit we derive characteristic ice sublimation speeds of 0.8 mm and 0.3 mm per Pluto year at the walls and floors, respectively, that are necessary to counteract the viscous relaxation of the $N_2$ ice and maintain the topography of the pit.

These values fall within the lower end of the range of sublimation rates determined for southern Sputnik Planitia in Pluto's current orbit (0 to 18 mm per Pluto year) by the global climate modeling of Bertrand et al. (2018), meaning that present sublimation rates may actually exceed the infilling rates due to viscous relaxation, and cause the pits to grow by several mm per Pluto year.

Moore et al. (2017) suggested that the wavelike patterns formed by swarms of large, elongate pits may develop in response to heterogeneous, anisotropic substrate properties, with pits nucleating from aligned structures in the surficial ices that would reflect stress fields associated with flow paths. Pit growth via sublimation erosion, exploiting existing structures (e.g. fractures and crevasses) formed at the surface of stiff, glacially-deforming $N_2$ ice, may progress at a different rate than sublimation erosion occurring at the surface of $N_2$ ice undergoing convection. Since dark floored pits are observed in expanses of non-cellular plains that are interstitial to convective cells (Figure 4d), the lateral transition from one spatial flow regime to the other can clearly be very abrupt, perhaps as little as a few kilometers.

Alternate formation mechanisms to sublimation erosion, including explosive/ejecta venting and structural collapse or collapse followed by sublimation erosion of the collapsed material on the resulting pit floor, are also worthy of discussion. We considered the explosion/ejecta venting pit origin hypothesis, but found it unlikely owing to several factors. First, the high-resolution LORRI images of the dark floored SP pits (e.g., Figures 2-4 above) do not show evidence of ejected material accumulating on pit rims or elsewhere near the pits. These images also reveal that adjoining pit walls are in many cases very thin compared to the pit dimensions: the inset in Fig. 4c that focuses on the cluster of especially large, elongate pits in southeastern SP shows that walls separating pits can be as thin as a few hundred meters for pits that typically reach several hundred meters across and many kilometers long. It is unlikely that such thin structures would remain intact if individual pits were formed in violent explosions or eruptions. The explosion/ejecta venting origin mechanism is also difficult to explain in light of the prevalent north-south orientation of the pits (see Figure 3A), which is more consistent with an insolation (i.e., sublimation) origin.

We also considered pit origin by structural collapse, but found that this mechanism is at odds with the fact that the collapsed material would consist of bright SP plains. The dark floors of the pits more naturally suggest an inert lag deposit remaining after the removal of volatiles by sublimation. For these reasons, we favor a sublimation origin for SP's pits. We cannot entirely rule out collapse followed by sublimation erosion of the collapsed material on the resulting pit floor, but this mechanism is less favored owing to the ubiquitously unbroken narrow wall structures seen at many pit-pit boundaries.

As to the nature and origin of the putative lag deposit in the SP pits, likely candidates are (i) cosmic-ray induced radiolysis of SP's ices and (ii) the solid tholin that forms as micron-size aggregate particles in Pluto's atmosphere, primarily produced by photolysis, and then precipitates to the surface (Gladstone et al. 2016, Cheng et al. 2017; Grundy et al. 2018; Protopapa et al. 2020) Supporting these concepts is the fact that SP's surface has a faint yellow-brown color, presumably resulting surface radiolysis and/or from precipitating haze particles.

## 8. Summary

The major conclusions we have reached in this study of the pit features on Sputnik Planitia are:

- ➢ In our sample of 317 pits, we find typical length/width ratios of 2-4 and major axis preferentially oriented approximately north-south.

- ➢ The floors of large pits in our sample have similar single scattering albedos and colors to dark material on crater rims and floors in crater windows on Burney basin, supporting (though not requiring) there being a potential common substrate beneath Pluto's ices as hypothesized above.

- ➢ The base of the three pits in our sample large enough to study with LEISA IR spectroscopy display both $CH_4$ and $N_2$ absorption features, as do the dark regions in crater windows on Burney basin.

- ➢ A sublimation erosion origin for the pits is supported over both the explosion/ejecta venting and structural collapse alternatives, with circumstances possibly involving fracturing and/or pit elongation via glacial flow, and the absence of convection being required for the growth of the large, dark floored pits.

- ➢ The missing mass represented by all of the pit volume on Sputnik Planitia most likely lies condensed elsewhere on Pluto's surface, relocated there by volatile transport, as opposed to loss by either escape to space or photochemical conversion.

- ➢ There is evidence for a widespread uniformity of exposed subsurface materials beneath volatile covered units across Pluto.

The dark floored pits of Sputnik Planitia are but one of many fascinating geomorphological features in Sputnik Planitia. Their further study, including their evident dynamics, and through modeling their growth by sublimation, are warranted. More work regarding the rheology of the volatile ices on Pluto's surface, and at those conditions and physio-chemical states, is also warranted.

## Acknowledgements

We thank the NASA New Horizons project for support. We also thank an anonymous referee for their helpful critique of this manuscript.